\documentclass[aps,prapplied,reprint,superscriptaddress,longbibliography]{revtex4-2}

\usepackage[T1]{fontenc}
\usepackage[utf8]{inputenc}
\usepackage{amsmath,amssymb,bm}
\usepackage{graphicx}
\usepackage{dcolumn}
\usepackage{xcolor}
\usepackage{hyperref}

\hypersetup{colorlinks=true,linkcolor=blue,citecolor=blue,urlcolor=blue}
\graphicspath{{./}}

\begin{document}

\title{Protecting Qubits from Purcell Decay via Permanent Dipoles}

\author{Alex Krasnok}
\email{akrasnok@fiu.edu}
\affiliation{Department of Electrical Engineering, Florida International University, Miami, Florida 33174, USA}
\affiliation{Knight Foundation School of Computing and Information Sciences, Florida International University, Miami, Florida 33174, USA}

\date{\today}

\begin{abstract}
Reading out a qubit often requires coupling it to a resonator, but that same resonator can also give the qubit an extra path to decay. Here, we study a way to reduce this loss using a built-in permanent electric dipole. The dipole shifts the cavity field in different directions for the qubit ground and excited states. This shift makes the relevant wave functions overlap less, which weakens the transverse qubit--cavity exchange that causes Purcell decay. In a simplified displaced rotating-wave model, this exchange vanishes at $\eta=\sqrt{2}$. In the full transverse model, this exact zero is lifted, but strong suppression remains at a larger dipole-induced displacement. Using dressed open-system decay rates, we find an operating point where the cavity-mediated decay is strongly reduced while the longitudinal readout signal remains finite. For the benchmark studied here, at fixed pointer separation, the normalized lifetime increases from $\kappa T_1=11.1$ to $47.3$, and the estimated single-shot readout error drops from $0.21$ to $0.07$. These results show that permanent electric dipoles can provide an internal, channel-selective form of Purcell protection.
\end{abstract}

\maketitle

\section{Introduction}

 {The fundamental tradeoff between coherent control and radiative loss defines a central design problem for open cavity quantum electrodynamics (QED).}  {Cavity QED operates by} coupling matter qubits to quantized electromagnetic modes, enabling precise coherent control, state transfer, and measurement \cite{Jaynes1963,Scully1997,Wallraff2004,Blais2021}.  {However, this interaction inevitably introduces} radiative loss: a qubit in a leaky cavity decays through the cavity mode, producing a Purcell  {decay} channel that limits fast readout, reset, and strong coupling \cite{Houck2008,Reed2010,Sete2015,Kockum2019}.

Transverse coupling  {($g_x$)} directly ties qubit control  {and} readout  {to} decay. In the Jaynes--Cummings limit, a qubit and a cavity swap one quantum  {of energy} through this coupling \cite{Jaynes1963,Scully1997}. If the cavity dissipates photons,  {this resonant} exchange exposes the qubit to Purcell decay. Circuit  {QED relies on} transverse coupling for strong coupling, dispersive readout, and entangling gates \cite{Wallraff2004,Blais2021}.  {While increasing this exchange accelerates} measurement and gates,  {it proportionally shortens the} qubit lifetime.

Longitudinal coupling  {($g_p$)} provides a  {distinct} measurement route  {that avoids resonant energy exchange}.  {This state-dependent force displaces} the cavity field in opposite directions for the qubit ground and excited states {, supporting quantum nondemolition (QND)} readout, geometric gates, and force-mediated interactions \cite{Kerman2013,Didier2015,Royer2017}. Superconducting circuits engineer this coupling through circuit asymmetry, parametric modulation, or driven frames \cite{Richer2016}. Quantum-dot experiments  {demonstrate analogous} longitudinal readout in a far-detuned, ultra-dispersive regime \cite{Harpt2025}.

 {Isolating the measurement signal from the decay mechanism remains an ongoing physical problem.} Standard methods  {manage decay using} external environmental Purcell filters to block leaky modes,  {leaving the internal matter-cavity interaction unchanged}.  {Questioning whether external environmental management is strictly necessary, we investigate whether} a single, built-in physical asymmetry can  {internally suppress the transverse matter-cavity exchange} that drives Purcell decay, while simultaneously preserving the longitudinal force  {required} for  {reliable measurement}.

 {Permanent dipole moments supply a mechanism for this internal suppression.}  {These static moments} arise naturally in systems lacking inversion symmetry, such as polar molecules, biased atoms, asymmetric quantum dots, moir'e interlayer excitons, and artificial superconducting circuits \cite{DeMille2002,Andre2006,Savenko2012,Baek2020,Yoshihara2017}. Because these systems exhibit unequal charge distributions in their functional states, they profoundly modify multiphoton transitions, fluorescence spectra, and open-system dynamics \cite{Meath1984,Kibis2009,Savenko2012,Macovei2015,Burgess2023}. Crucially, this asymmetry introduces wave-function overlap zeros {---a phenomenon analogous to} the Franck--Condon blockade  {in molecular transport}, where displaced oscillator wave functions strongly suppress state transitions \cite{Koch2005}. Similar overlap zeros  {collapse} Rabi oscillations \cite{Baranov2016} and alter polariton photochemistry \cite{Mandal2020}.

\begin{figure*}[t]
\centering
\includegraphics[width=0.98\textwidth]{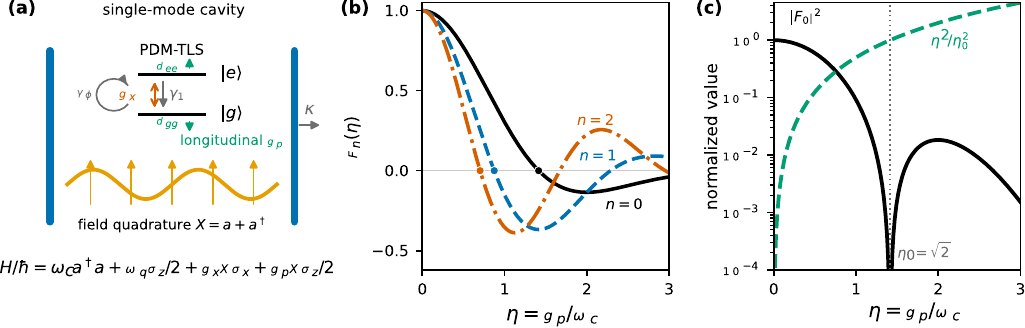}
\caption{Permanent-dipole exchange suppression. (a) The model contains a transverse exchange coupling $g_x$ and a longitudinal permanent-dipole coupling $g_p$. (b)  {Franck--Condon} factors $F_n(\eta)$ set the exchange strength in the first three excitation manifolds;  {markers} denote the first zeros. (c)  {In the lowest manifold, the displaced-RWA transverse-exchange proxy collapses near $\eta_0=\sqrt{2}$, while the longitudinal readout-force proxy remains finite. This proxy is not the full-transverse cavity-loss rate.}  {Panels (b) and (c) plot analytic} evaluations of Eq.~(\ref{eq:collapse_factor}) and the proxies $|F_0|^2$ and $\eta^2$.}
\label{fig:fig1}
\end{figure*}

 {Applying this overlap-zero physics, we show that a permanent dipole generates the longitudinal coupling ($g_p$) and suppresses the transverse exchange channel ($g_x$) that contributes to Purcell decay. Using an open-cavity model, we derive the displaced-Fock exchange factor $F_n(\eta)$ and identify the first displaced-RWA exchange zero. We then compare this analytical zero with a full-transverse dressed calculation. The full-transverse model lifts the exact zero at $\eta_0=\sqrt{2}$, but it retains strong cavity-channel suppression at a larger operating point. Finally, we evaluate a reduced readout benchmark at fixed pointer separation to separate lifetime gain from phase-space distinguishability.}

\section{Results and discussion}

\subsection{Permanent-dipole cavity-QED model}

We model one cavity mode coupled to an asymmetric two-level system. Let $a$ be the cavity annihilation operator, $\omega_c$ the cavity frequency, and $\omega_q$ the qubit transition frequency. The Pauli operators $\sigma_x$ and $\sigma_z$ act on the qubit states $|g\rangle$ and $|e\rangle$. Projecting the electric-dipole interaction onto this two-state subspace and retaining one cavity quadrature gives
\begin{equation}
\frac{H}{\hbar}=\omega_c a^\dagger a+\frac{\omega_q}{2}\sigma_z
+g_x(a+a^\dagger)\sigma_x+\frac{g_p}{2}(a+a^\dagger)\sigma_z .
\label{eq:hamiltonian}
\end{equation}
Equation~(\ref{eq:hamiltonian}) is an effective two-level Hamiltonian; Appendix~\ref{app:gauge} states the gauge and truncation limits that a platform-specific derivation must satisfy. The transverse coupling $g_x=d_{eg}E_{\rm vac}/\hbar$  {arises} from the transition dipole $d_{eg}$ and the vacuum electric field $E_{\rm vac}$. The permanent-dipole coupling $g_p=(d_{ee}-d_{gg})E_{\rm vac}/\hbar$  {arises} from the difference between diagonal dipoles in the two qubit states. We use the dimensionless longitudinal coupling
\begin{equation}
\eta=\frac{g_p}{\omega_c}  {.}
\label{eq:eta_def}
\end{equation}
 {This} parameter  {normalizes} the static conditional displacement to the cavity frequency and fixes the displaced-oscillator coordinate used in Appendix~\ref{app:derivation}.

Figure~\ref{fig:fig1}(a) identifies the two competing couplings in Eq.~(\ref{eq:hamiltonian}). The $g_x$ term is the transverse photon-exchange channel that allows the qubit and cavity to trade an excitation. The $g_p$ term is the longitudinal coupling that pushes the cavity field in opposite directions for the two qubit states. We use ``longitudinal cavity force'' and ``state-dependent force'' for this conditional displacement.

\begin{table*}[t]
\caption{ {Platform} requirements and dominant risks for reaching the first displaced-overlap zero.}
\label{tab:platforms}
\small
\setlength{\tabcolsep}{3pt}
\begin{center}
\begin{tabular}{lll}
\hline
\begin{minipage}[t]{0.18\textwidth}\textbf{Platform}\end{minipage} &
\begin{minipage}[t]{0.43\textwidth}\textbf{Route to large $\eta$}\end{minipage} &
\begin{minipage}[t]{0.30\textwidth}\textbf{Main risk}\end{minipage} \\
\hline
\begin{minipage}[t]{0.18\textwidth}Polar molecules\end{minipage} &
\begin{minipage}[t]{0.43\textwidth} {Coupling large} permanent electric dipoles to microwave resonators with high vacuum fields \cite{DeMille2002,Andre2006,Park2017,Gregory2024}.\end{minipage} &
\begin{minipage}[t]{0.30\textwidth} {Electric-field noise, motional} averaging, and molecule-resonator mode overlap.\end{minipage} \\
\begin{minipage}[t]{0.18\textwidth}Asymmetric quantum dots\end{minipage} &
\begin{minipage}[t]{0.43\textwidth} {Engineering large} charge displacement and tunable longitudinal response in semiconductor resonators \cite{Harpt2025}.\end{minipage} &
\begin{minipage}[t]{0.30\textwidth} {Charge noise, leakage} to higher orbital or valley states, and drive-induced transitions.\end{minipage} \\
\begin{minipage}[t]{0.18\textwidth}Moir\'e excitons and solid-state emitters\end{minipage} &
\begin{minipage}[t]{0.43\textwidth} {Exploiting broken} inversion symmetry and large DC Stark response \cite{Baek2020,Savenko2012}.\end{minipage} &
\begin{minipage}[t]{0.30\textwidth} {Phonons, disorder, optical-frequency scaling}, and radiative loss outside the target mode.\end{minipage} \\
\begin{minipage}[t]{0.18\textwidth}Artificial superconducting circuits\end{minipage} &
\begin{minipage}[t]{0.43\textwidth} {Engineering effective} diagonal dipoles and deep-strong oscillator coupling \cite{Yoshihara2017,Kockum2019}.\end{minipage} &
\begin{minipage}[t]{0.30\textwidth} {Gauge consistency, counter-rotating terms}, and flux or charge noise.\end{minipage} \\
\hline
\end{tabular}
\end{center}
\end{table*}

The permanent dipole conditionally displaces the cavity oscillator and changes the exchange matrix element. Appendix~\ref{app:derivation} derives this result by applying a conditional displacement to the rotating-wave Hamiltonian, retaining only the energy-conserving transverse terms $a\sigma_+$ and $a^\dagger\sigma_-$. In the $n$th excitation manifold, the transverse matrix element in angular-frequency units is
\begin{equation}
M_n=g_x\sqrt{n+1}\,F_n(\eta),
\label{eq:matrix_element}
\end{equation}
where the dimensionless Franck--Condon factor is
\begin{equation}
F_n(\eta)=\frac{1}{2}e^{-\eta^2/2}\left[L_n(\eta^2)+L_{n+1}(\eta^2)\right].
\label{eq:collapse_factor}
\end{equation}
Equation~(\ref{eq:collapse_factor}) follows from the displaced-Fock overlap identity used in Appendix~\ref{app:derivation}; it is exact within the displaced rotating-wave approximation  {(RWA)}, not within the full transverse Hamiltonian. Here $L_n$ is an ordinary Laguerre polynomial, and $\sqrt{n+1}$ is the Jaynes--Cummings enhancement. For the lowest manifold, $F_0(\eta)=e^{-\eta^2/2}(1-\eta^2/2)$, so the first displaced-RWA exchange zero occurs at
\begin{equation}
\eta_0=\sqrt{2}.
\label{eq:eta0}
\end{equation}

 {This zero is a displaced-RWA zero, not an exact zero of the physical transverse operator in Eq.~(\ref{eq:hamiltonian}). For the same displaced states, the lowest-manifold matrix element of the full lowering part is}
\begin{equation}
{
\begin{aligned}
\frac{\mathcal M_0^{\rm FT}}{g_x}
&={}_d\!\langle g,1|(a+a^\dagger)\sigma_-|e,0\rangle_d \\
&=\langle 1|(a+a^\dagger+\eta)D(-\eta)|0\rangle \\
&=e^{-\eta^2/2}.
\end{aligned}}
\label{eq:full_transverse_m0}
\end{equation}
 {The displaced-RWA contribution $a^\dagger\sigma_-$ gives $e^{-\eta^2/2}(1-\eta^2/2)$, whereas the omitted $a\sigma_-$ term contributes $e^{-\eta^2/2}\eta^2/2$. Their sum removes the exact zero at $\eta=\sqrt{2}$. This identity explains the lifted minimum in the full-transverse rate calculation below and motivates using a larger full-transverse operating point in Fig.~\ref{fig:fig3}.}

To demonstrate manifold selectivity, Fig.~\ref{fig:fig1}(b) plots the analytical factors $F_n(\eta)$ for $n=0,1,2$. The first zeros occur at $\eta=1.414$, $0.874$, and $0.702$, respectively.  {Because a} drive that populates higher manifolds samples different exchange zeros,  {this mechanism does not offer broadband protection. We analyze the lowest-manifold displaced-RWA exchange zero,} not an exact zero of the full physical Hamiltonian at all photon numbers.

The hardware target follows from Eqs.~(\ref{eq:eta_def}) and (\ref{eq:eta0}) using the standard electric-dipole cavity coupling convention in Eq.~(\ref{eq:hamiltonian}) \cite{Andre2006,Blais2021}. To suppress the lowest-manifold exchange channel, the required diagonal dipole difference is
\begin{equation}
\Delta d_{\rm req}=\frac{\sqrt{2}\hbar\omega_c}{E_{\rm vac}}
\simeq 1.4~{\rm D}\left(\frac{\omega_c/2\pi}{5~{\rm GHz}}\right)
\left(\frac{10^6~{\rm V/m}}{E_{\rm vac}}\right),
\label{eq:d_req}
\end{equation}
where ${\rm D}$ denotes the Debye. Equation~(\ref{eq:d_req})  {links} the overlap-zero condition to the actual vacuum field of a device. Table~\ref{tab:platforms} summarizes the platform requirements and dominant noise risks.

To show that the overlap zero does not remove the longitudinal force, Fig.~\ref{fig:fig1}(c) compares two analytical lowest-manifold proxies:  {the displaced-RWA transverse-exchange proxy} $|F_0(\eta)|^2$ and the longitudinal readout-force proxy $\eta^2$.  {The displaced-RWA proxy} vanishes at $\eta_0=\sqrt{2}$, whereas the longitudinal proxy remains finite and grows with $\eta$.  {This comparison identifies} the parameter point where the targeted exchange matrix element is suppressed while the state-dependent cavity displacement remains available  {without requiring a full master-equation decay rate calculation}.

\subsection{Dressed open-system rates}

 {Because a} permanent dipole displaces and hybridizes the bare photon and qubit states,  {open-system calculations require dressed states.} We diagonalize the displaced rotating-wave Hamiltonian in a truncated Fock basis and label the eigenstates $|i\rangle$ by increasing energy $E_i$. In the secular, zero-temperature Markov approximation, the population decay rate from state $|i\rangle$ is \cite{Beaudoin2011,Kockum2019}
\begin{equation}
\Gamma_i=\sum_{\mu}\sum_{E_j<E_i}\gamma_\mu
\left|\langle j|S_\mu|i\rangle\right|^2 .
\label{eq:dressed_rate}
\end{equation}
Equation~(\ref{eq:dressed_rate}) is the rate form of a Lindblad master equation written in the system eigenbasis {.} Appendix~\ref{app:numerics}  {details} the corresponding positive-frequency dressed-bath construction {, Hamiltonians, branch-labeling rules, truncation cutoff, and parameter values}. The channel index $\mu$ includes cavity leakage ($S_c=a$, rate $\gamma_c=\kappa$), intrinsic transverse relaxation ($S_x=\sigma_x$, rate $\gamma_x=\gamma_1$), and longitudinal dephasing ($S_z=\sigma_z$, rate $\gamma_z=\gamma_\phi$).

\begin{figure*}[t]
\centering
\includegraphics[width=0.98\textwidth]{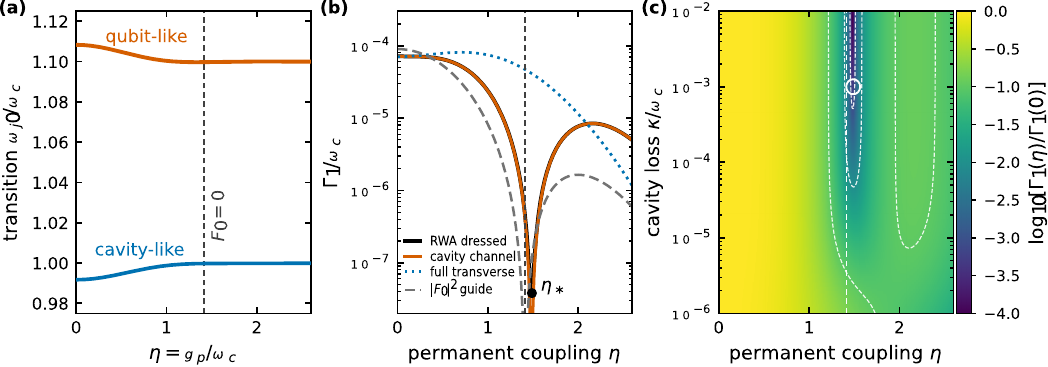}
\caption{Dressed open-system rates. (a) Qubit-like and cavity-like transition frequencies in the displaced rotating-wave model versus longitudinal coupling $\eta$. (b) Dressed qubit-like decay compared  {to} the full-transverse calculation and the analytical $|F_0(\eta)|^2$ guide. (c) Rate reduction versus $\eta$ and normalized cavity decay $\kappa/\omega_c$.  {Panel (a) plots Hamiltonian diagonalization; panels (b) and (c) plot} numerical dressed-state evaluations of the zero-temperature secular Lindblad rates  {from} Eq.~(\ref{eq:dressed_rate}).}
\label{fig:fig2}
\end{figure*}

 {We numerically evaluate Eq.~(\ref{eq:dressed_rate}) to determine if the analytical exchange zero suppresses an open-system decay channel.}  {We use parameters} $g_x/\omega_c=0.030$, detuning $\Delta/\omega_c=0.10$, $\kappa/\omega_c=10^{-3}$, $\gamma_1/\omega_c=10^{-7}$, and $\gamma_\phi/\omega_c=3\times10^{-7}$. The calculation diagonalizes the Hamiltonian at each $\eta$, identifies qubit-like and cavity-like dressed branches by displaced-state overlap, and sums the squared matrix elements of $a$, $\sigma_x$, and $\sigma_z$ to lower-energy dressed states. Figure~\ref{fig:fig2}(a) tracks  {these} branches by their overlaps with displaced reference states {; they} remain distinct near $\eta_0$ because finite detuning prevents resonant hybridization. Figure~\ref{fig:fig2}(b) compares the dressed decay rate  {to} the analytical $|F_0(\eta)|^2$ guide {. The} displaced-RWA rate follows the guide away from the minimum and then saturates at a floor set by intrinsic relaxation, longitudinal-noise-induced transitions, and finite detuning.

 {These numerical data quantify the displaced-RWA suppression scale. At the grid point nearest $\eta_0$, the total displaced-RWA rate drops to $5.5\times10^{-3}$ of its unprotected $\eta=0$ value. The displaced-RWA minimum occurs at $\eta=1.489$ and reaches $5.3\times10^{-4}$ of the unprotected rate. The rate remains below $1\%$ of its unprotected value for $1.39\leq \eta \leq 1.58$. This interval defines an operational tolerance window for the parameters in Fig.~\ref{fig:fig2}(b).}

The full-transverse calculation in Fig.~\ref{fig:fig2}(b) tests the same parameter sweep after restoring counter-rotating transverse terms. At the grid point nearest $\eta_0$, the total full-transverse rate  {drops} only to $0.66$ of its $\eta=0$ value. Within the plotted range, the minimum full-transverse rate occurs at the upper boundary, $\eta=2.60$, and reaches $1.6\times10^{-2}$ of the uncoupled value. This deviation is expected because Eq.~(\ref{eq:collapse_factor}) relies on the displaced-RWA, whereas the full Hamiltonian adds counter-rotating terms and gauge-sensitive truncation effects typical of ultrastrong-coupling models \cite{Kockum2019,Stokes2019,DiStefano2019}.  {Therefore, this mechanism provides} channel-selective Purcell suppression near a displaced-Fock zero, not exact decoupling from all radiative channels.  {We therefore distinguish the displaced-RWA zero $\eta_0=\sqrt{2}$ from the full-transverse benchmark point $\eta_{\rm FT}=2.60$, which is a representative high-suppression point within the parameter sweep rather than a universal root.}

 {To test the stability of the protected point against changes in cavity loss, Fig.~\ref{fig:fig2}(c) maps} $\log_{10}[\Gamma_1(\eta)/\Gamma_1(0)]$ for $10^{-6}\leq\kappa/\omega_c\leq10^{-2}$ using the displaced rotating-wave model. The map reuses the dressed matrix elements from the displaced-RWA diagonalization and rescales only the cavity-loss prefactor $\kappa$. The low-rate valley aligns with the overlap zero and broadens when residual non-cavity rates set the decay floor. The practical operating point is therefore a tolerance window defined by competition between the suppressed cavity-mediated channel and the  {residual} decay channels.

\subsection{Longitudinal readout benchmark}

We use a readout benchmark to isolate the qubit lifetime from pointer separation {---}the distinguishability of the cavity field states. In the weak-exchange, detuned regime $\Delta=\omega_q-\omega_c$, the cavity-mediated contribution to population decay is approximated by the standard Purcell expression with the displaced-RWA exchange matrix element inserted \cite{Blais2021,Houck2008,Sete2015}:
\begin{equation}
\Gamma_{1,{\rm cav}}(\eta)\simeq \kappa
\frac{|g_xF_0(\eta)|^2}{\Delta^2+(\kappa/2)^2} .
\label{eq:purcell}
\end{equation}
Equation~(\ref{eq:purcell})  {serves as} an analytical guide, not the primary dressed-rate calculation. It shows why the permanent dipole suppresses the Purcell channel in the displaced-RWA limit: the cavity-mediated decay scales directly with $|F_0(\eta)|^2$.

\begin{figure*}[t]
\centering
\includegraphics[width=0.98\textwidth]{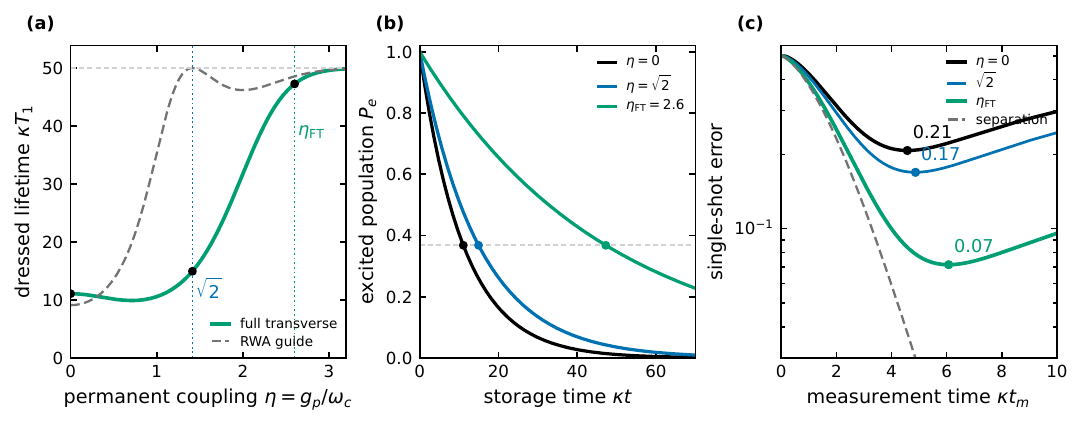}
\caption{ {Storage and longitudinal-readout benchmark using full-transverse dressed rates. (a) Normalized dressed lifetime $\kappa T_1$ versus permanent-dipole displacement. The dashed curve is the displaced-RWA analytical guide, while the solid curve uses the full-transverse dressed cavity-loss rate in Eq.~(\ref{eq:ft_benchmark_rate}). Vertical dotted lines mark $\eta_0=\sqrt{2}$ and $\eta_{\rm FT}=2.60$. (b) Storage envelopes for $\eta=0$, $\eta=\sqrt{2}$, and $\eta_{\rm FT}$. (c) Estimated single-shot readout error at matched steady-state pointer separation.}}
\label{fig:fig3}
\end{figure*}

 {For the benchmark in Fig.~\ref{fig:fig3}, we instead use the full-transverse dressed cavity-loss matrix element. Let $|q(\eta)\rangle$ denote the qubit-like dressed eigenstate and let $|j(\eta)\rangle$ denote lower-energy dressed states. The cavity-mediated component is}
\begin{equation}
{
\begin{aligned}
\Gamma^{\rm FT}_{1,{\rm cav}}(\eta)
&=\kappa\sum_{E_j<E_q}\left|\langle j(\eta)|a|q(\eta)\rangle\right|^2,\\
\Gamma^{\rm bench}_{1}(\eta)
&=\gamma_{1,{\rm floor}}+\Gamma^{\rm FT}_{1,{\rm cav}}(\eta).
\end{aligned}}
\label{eq:ft_benchmark_rate}
\end{equation}
 {The fixed floor $\gamma_{1,{\rm floor}}$ represents non-cavity relaxation and prevents the benchmark from claiming that permanent-dipole displacement suppresses unrelated material loss.}

A readout measurement requires converting the static longitudinal coupling into a resonant rotating-frame force. The benchmark assumes that the diagonal dipole difference, the cavity field, or an external bias is modulated to generate a longitudinal force with amplitude $\tilde g_z$. Appendix~\ref{app:readout} derives the rotating-frame equations from the minimal parametrization
\begin{equation}
g_p(t)=g_p^{(0)}+\delta g_p\cos(\omega_c t+\phi),
\label{eq:gp_mod}
\end{equation}
where $g_p^{(0)}$ sets the static displaced-overlap physics and the resonant component generates the readout force after the  {rotating-wave approximation}. Assuming coherent pointer states and a common cavity drive amplitude $\epsilon_d$, Appendix~\ref{app:readout}  {yields}
\begin{equation}
\dot{\alpha}_{e/g}=-\frac{\kappa}{2}\alpha_{e/g}-i\epsilon_d\pm i\tilde g_z.
\label{eq:pointer}
\end{equation}
The measured signal is the dynamic separation $\Delta\alpha(t)=\alpha_e(t)-\alpha_g(t)$ {. The} common drive shifts both pointers equally and does not increase distinguishability.

We estimate the single-shot readout error $\epsilon_{\rm tot}$ at measurement time $t_m$ as
\begin{equation}
\begin{aligned}
\epsilon_{\rm tot}(t_m)
&=\frac{1}{2}\operatorname{erfc}\left[\sqrt{\frac{\eta_{\rm det}\kappa}{2}
\int_0^{t_m}|\Delta\alpha(t)|^2dt}\right] \\
&\quad+\frac{1}{2}\left(1-e^{-t_m/T_1}\right),
\end{aligned}
\label{eq:error}
\end{equation}
where $\eta_{\rm det}$ is the detection efficiency and $T_1$ is the total dressed relaxation time.  {In Fig.~\ref{fig:fig3}, $T_1=[\Gamma^{\rm bench}_{1}(\eta)]^{-1}$ from Eq.~(\ref{eq:ft_benchmark_rate}).}  {The first term represents} the Gaussian separation error {, utilizing the pointer-separation integral from Appendix~\ref{app:readout}.}  {The second term} estimates errors from unresolved qubit relaxation during the measurement. By comparing devices at fixed pointer separation, this benchmark isolates the penalty from  {a} short $T_1$. It omits measurement-induced transitions, stochastic trajectory backaction, finite thermal occupation, resonator nonlinearity, drive-power limits, and leakage.

 {Figure~\ref{fig:fig3}(a) plots the lifetime enhancement obtained from Eq.~(\ref{eq:ft_benchmark_rate}) with $\Delta/\omega_c=0.10$, $g_x/\omega_c=0.030$, $\kappa/\omega_c=10^{-3}$, and $\gamma_{1,{\rm floor}}/\omega_c=2\times10^{-5}$. The displaced-RWA guide reaches its first zero at $\eta_0=\sqrt{2}$, but the full-transverse dressed lifetime is only modestly improved there. The strong full-transverse effect appears at larger displacement. At $\eta_{\rm FT}=2.60$, the dressed cavity-mediated rate is reduced close to the non-cavity floor, giving $\kappa T_1=47.3$ compared with $11.1$ at $\eta=0$ and $14.9$ at $\eta=\sqrt{2}$.}

 {Figure~\ref{fig:fig3}(b) uses these dressed lifetimes to plot storage envelopes. Figure~\ref{fig:fig3}(c) evaluates Eq.~(\ref{eq:error}) at matched steady-state pointer separation $|\Delta\alpha_{\rm ss}|=2$ and detection efficiency $\eta_{\rm det}=0.40$. The minimum illustrative readout error changes from $0.207$ for $\eta=0$ to $0.169$ at $\eta=\sqrt{2}$ and $0.071$ at $\eta_{\rm FT}$. Thus the full-transverse benchmark demonstrates strong Purcell-channel suppression, but the useful operating point is shifted away from the displaced-RWA zero.}

 {This benchmark demonstrates a mechanism-level lifetime gain, not a complete readout architecture. A device-level claim requires a driven-dissipative master equation or quantum-trajectory simulation at fixed measurement power or fixed intracavity photon number. Such a model must include the physical modulation that generates $\tilde g_z$, modulation-induced noise, and unwanted transitions to higher oscillator or material levels.}

\subsection{On the protection mechanism}

 {This} mechanism suppresses a transition matrix element {, rather than} intrinsic relaxation or dephasing. It  {operates independently of} exceptional-point sensitivity and does not require eigenvector coalescence.  {Its} useful regime is limited to the parameter window where the suppressed cavity-mediated transverse channel falls below residual non-cavity channels.  {In the displaced-RWA model this suppression can appear as an exact overlap zero; in the full-transverse model it becomes a shifted Franck--Condon suppression of the dressed cavity-loss matrix element.}

Established Purcell-protection methods {---such as d}etuning, impedance shaping, bandpass Purcell filters, and reset/readout filters {---}engineer the electromagnetic environment  {to} reduce the local density of leaky modes  {coupled to} the qubit \cite{Houck2008,Reed2010,Sete2015}.  {In contrast, our} mechanism reduces the internal qubit-cavity matrix element within a selected excitation manifold. It  {complements, rather than replaces,} external filters,  {device-level impedance models, and noise budgets.}  {Table~\ref{tab:platforms} should be read as a feasibility map, not a platform proof.}

The primary hardware limitation is electric-field noise. Increasing the diagonal dipole difference increases $g_p$, but  {simultaneously elevates} sensitivity to low-frequency electric-field fluctuations. The critical metric is the ratio between the suppressed Purcell rate and the added pure dephasing or leakage caused by  {this} microscopic asymmetry.  {Furthermore, our} model assumes that the transverse dipole remains available while the diagonal dipole difference is tuned.

Permanent dipoles  {provide a tunable design knob to suppress} a specific transverse exchange matrix element while retaining a longitudinal force. A symmetric transition-dipole-only emitter cannot provide this separation.  {Implementing this mechanism in a hardware architecture requires validation} with platform-specific bath spectra, gauge-consistent Hamiltonians, and noise budgets.

\section{Conclusion}

 {A permanent electric dipole can suppress the cavity-mediated Purcell channel by displacing the oscillator wave functions associated with the two qubit states. In the displaced-RWA model this suppression appears as an exact lowest-manifold exchange zero at $\eta=\sqrt{2}$. In the full-transverse Hamiltonian, counter-rotating transverse terms lift that zero and shift the useful protection point to larger $\eta$. For the representative full-transverse dressed calculation used here, $\eta_{\rm FT}=2.60$ suppresses the dressed cavity-mediated decay close to the non-cavity relaxation floor, increasing $\kappa T_1$ from $11.1$ to $47.3$ at fixed pointer separation. The mechanism therefore provides channel-selective Purcell suppression rather than exact decoupling from all radiative and material baths. Implementations must validate the effective Hamiltonian, the readout modulation, the bath spectra, and the added electric-field-noise budget.}

\begin{acknowledgments}
The author acknowledges financial support from the U.S. Department of Energy (DoE) and the U.S. Air Force Office of Scientific Research (AFOSR).
\end{acknowledgments}

\appendix

\section{Displaced-frame derivation of the exchange factor}
\label{app:derivation}

We derive Eq.~(\ref{eq:collapse_factor}) from the displaced  {rotating-wave approximation (RWA)} Hamiltonian. This approximation retains only the energy-conserving exchange terms $a\sigma_+$ and $a^\dagger\sigma_-$ in the transverse coupling of Eq.~(\ref{eq:hamiltonian}):
\begin{equation}
\frac{H_{\rm RWA}}{\hbar}=\omega_c a^\dagger a+\frac{\omega_q}{2}\sigma_z
+g_x(a\sigma_+ +a^\dagger\sigma_-)+\frac{g_p}{2}(a+a^\dagger)\sigma_z .
\label{eq:hrwa_app}
\end{equation}
The conditional displacement operator
\begin{equation}
U=\exp\left[-\frac{\eta}{2}(a^\dagger-a)\sigma_z\right]
\label{eq:displacement_unitary}
\end{equation}
diagonalizes the longitudinal term up to a constant energy shift. Here $D(\alpha)=\exp(\alpha a^\dagger-\alpha^*a)$ is the oscillator displacement operator; the displacements used  {here} are real. The transformed oscillator states are
\begin{equation}
|e,n\rangle_d=D(-\eta/2)|n\rangle|e\rangle,\qquad
|g,n\rangle_d=D(+\eta/2)|n\rangle|g\rangle .
\label{eq:displaced_states}
\end{equation}
The transverse exchange energy matrix element $M_n$ between adjacent excitation manifolds is
\begin{equation}
M_n=\hbar g_x\,{}_d\!\langle g,n+1|a^\dagger\sigma_-+a\sigma_+|e,n\rangle_d .
\label{eq:mn_def}
\end{equation}
Dividing this matrix element by the bare Jaynes--Cummings scaling factor $\hbar g_x\sqrt{n+1}$ isolates the dimensionless overlap factor $F_n(\eta)$ in Eq.~(\ref{eq:collapse_factor}). For the lowest manifold ($n=0$), the bare scaling factor is unity, so $M_0=\hbar g_xF_0(\eta)$.  {Consequently, the} exchange zero at $\eta_0=\sqrt{2}$ is exact only within the displaced  {RWA}.

 {The full transverse operator adds the lowering counter-rotating term $a\sigma_-$. For $n=0$,}
\begin{equation}
{
\begin{aligned}
\frac{\mathcal M_0^{\rm FT}}{g_x}
&={}_d\!\langle g,1|(a+a^\dagger)\sigma_-|e,0\rangle_d \\
&=\langle 1|D(-\eta/2)(a+a^\dagger)D(-\eta/2)|0\rangle \\
&=\langle 1|(a+a^\dagger+\eta)D(-\eta)|0\rangle \\
&=e^{-\eta^2/2}.
\end{aligned}}
\label{eq:app_full_transverse_m0}
\end{equation}
 {The first term $a^\dagger\sigma_-$ alone gives $e^{-\eta^2/2}(1-\eta^2/2)$, whereas $a\sigma_-$ gives $e^{-\eta^2/2}\eta^2/2$. Therefore the physical lowest-manifold full-transverse matrix element has no zero at $\eta=\sqrt{2}$. Higher-manifold transitions and physical bath interactions can further shift the minimum.}

\section{Gauge origin and two-level limits}
\label{app:gauge}

Predictive platform models must derive the effective two-level Hamiltonian  {[Eq.~(\ref{eq:hamiltonian})]} from a gauge-fixed Pauli--Fierz Hamiltonian or circuit Lagrangian.  {This requires truncating} to two material states after identifying the physical field quadrature coupled to the external bath. Molecular  {cavity-QED} models must retain the self-polarization term, and circuit models must retain the analogous capacitive or inductive renormalization before truncation. Dropping these terms can shift frequencies, change matrix elements, and produce gauge-dependent predictions.  {The present work therefore treats Eq.~(\ref{eq:hamiltonian}) as a broad effective Hamiltonian for the overlap-suppression mechanism, not as a complete derivation for a specific material or circuit platform.}

The coupling rates $g_x$ and $g_p$ are dressed, device-level parameters. The transverse coupling $g_x$ originates from the transition dipole between the selected states, and the longitudinal coupling $g_p$ arises from the difference between their diagonal dipoles. A platform calculation must also test whether the required microscopic asymmetry introduces low-frequency electric-field noise, leakage to nearby material states, or unaccounted radiative decay channels.

The two-level approximation is strongest when all other material transitions are far detuned from the cavity and readout modulation sidebands.  {Although the} oscillator displacement near $\eta_0=\sqrt{2}$ is substantial, the photon-number cutoff used in Fig.~\ref{fig:fig2} converges for the plotted undriven rates. Oscillator-basis convergence does not prove the material two-level approximation; it only shows that the truncated oscillator basis spans the effective Hamiltonian analyzed here.

\section{Numerical diagonalization and dressed-rate calculation}
\label{app:numerics}

We diagonalize the Hamiltonian in the product basis $|n\rangle\otimes\{|g\rangle,|e\rangle\}$ with  {a} photon-number cutoff $0\leq n<n_{\rm cut}$. All figures use $n_{\rm cut}=30$  {and employ} dense Hermitian diagonalization at each value of $\eta$. The largest static displacement in Fig.~\ref{fig:fig2} is $|\eta|/2=1.3$, so the displaced-state photon-number tail lies below this cutoff. Device-specific simulations should  {report convergence versus $n_{\rm cut}$ because stronger drives or higher excitation manifolds populate larger photon numbers.}

The displaced  {RWA} calculation diagonalizes Eq.~(\ref{eq:hrwa_app}), whereas the full-transverse model diagonalizes Eq.~(\ref{eq:hamiltonian})  {retaining} the full $g_x(a+a^\dagger)\sigma_x$ interaction. At each $\eta$, we assign the qubit-like branch by maximum overlap with $D(-\eta/2)|0\rangle|e\rangle$ within the single-excitation frequency window. We assign the cavity-like branch by maximum overlap with $D(+\eta/2)|1\rangle|g\rangle$.

We evaluate the local Markov decay rate used in the main figures with Eq.~(\ref{eq:dressed_rate}) {, serving as a numerical solution of the secular Lindblad rate equations in the dressed basis initialized in the qubit-like dressed state.}  {For systems requiring frequency-dependent baths or finite temperature, the population-transfer rate must include both emission and absorption terms:}
\begin{equation}
{
\begin{aligned}
\Gamma_i=&\sum_{\mu}\sum_{E_j<E_i}J_\mu(\omega_{ij})[n_\mu(\omega_{ij})+1]
\left|\langle j|S_\mu|i\rangle\right|^2 \\
&+\sum_{\mu}\sum_{E_j>E_i}J_\mu(\omega_{ji})n_\mu(\omega_{ji})
\left|\langle j|S_\mu|i\rangle\right|^2,
\end{aligned}}
\label{eq:general_rate}
\end{equation}
 {where $\omega_{ij}=(E_i-E_j)/\hbar>0$ for the downward term, $\omega_{ji}=(E_j-E_i)/\hbar>0$ for the upward term, $J_\mu$ is the bath spectral density, and $n_\mu$ is the thermal occupation. The zero-temperature expression in Eq.~(\ref{eq:dressed_rate}) is recovered by setting $n_\mu=0$ and using flat spectra.} Equivalently, one can construct the positive-frequency bath-coupled operator
\begin{equation}
S_\mu^{(+)}=\sum_{E_i>E_j}|j\rangle\langle j|S_\mu|i\rangle\langle i| .
\label{eq:positive_frequency}
\end{equation}
This construction avoids applying bare local jump operators outside their validity range. It is essential in ultrastrong-coupling regimes and in gauge-sensitive two-level truncations \cite{Beaudoin2011,Kockum2019,Stokes2019,DiStefano2019}.

For reproducibility, we list the parameters used across the figures. Figure~\ref{fig:fig2} uses $g_x/\omega_c=0.030$, $\Delta/\omega_c=0.10$, $\kappa/\omega_c=10^{-3}$, $\gamma_1/\omega_c=10^{-7}$, $\gamma_\phi/\omega_c=3\times10^{-7}$, $0\leq\eta\leq2.6$, and $10^{-6}\leq\kappa/\omega_c\leq10^{-2}$ in the two-dimensional rate map.  {Figure~\ref{fig:fig3} uses the full-transverse Hamiltonian with $n_{\rm cut}=40$, $0\leq\eta\leq3.2$, $\eta_{\rm FT}=2.60$, $\gamma_{1,{\rm floor}}/\omega_c=2\times10^{-5}$, $|\Delta\alpha_{\rm ss}|=2$, and $\eta_{\rm det}=0.40$. It keeps $g_x/\omega_c=0.030$, $\Delta/\omega_c=0.10$, and $\kappa/\omega_c=10^{-3}$. The plotted storage and discrimination-error curves use Eq.~(\ref{eq:ft_benchmark_rate}) for population loss and omit pure dephasing.}

\section{Readout model and benchmark limits}
\label{app:readout}

We derive Eq.~(\ref{eq:pointer}) by assuming a resonant longitudinal force in the cavity rotating frame. Starting from the modulated coupling in Eq.~(\ref{eq:gp_mod}), the  {RWA} gives the resonant longitudinal-readout Hamiltonian
\begin{equation}
\frac{H_{\rm ro}}{\hbar}=\tilde g_z(a+a^\dagger)\sigma_z+\epsilon_d(a+a^\dagger),
\label{eq:hro}
\end{equation}
where $\tilde g_z$ is the effective readout-force amplitude after absorbing modulation phases and numerical factors {.} Integrating Eq.~(\ref{eq:pointer}) with the empty-cavity initial condition $\alpha_e(0)=\alpha_g(0)=0$ gives the time-dependent pointer separation
\begin{equation}
\Delta\alpha(t)=\frac{4i\tilde g_z}{\kappa}\left(1-e^{-\kappa t/2}\right).
\label{eq:pointer_solution}
\end{equation}
Evaluating the squared magnitude over time gives the separation integral required for Eq.~(\ref{eq:error}):
\begin{equation}
\begin{aligned}
\int_0^{t_m}|\Delta\alpha(t)|^2dt
&=|\Delta\alpha_{\rm ss}|^2
\bigg[t_m-\frac{4}{\kappa}\left(1-e^{-\kappa t_m/2}\right)\\
&\quad+\frac{1}{\kappa}\left(1-e^{-\kappa t_m}\right)\bigg],
\end{aligned}
\label{eq:sep_integral}
\end{equation}
with the steady-state separation $|\Delta\alpha_{\rm ss}|=4|\tilde g_z|/\kappa$.  {For the Fig.~\ref{fig:fig3} benchmark, $|\Delta\alpha_{\rm ss}|=2$, so $|\tilde g_z|/\kappa=0.50$.}

This benchmark isolates  {the performance improvement gained by suppressing the cavity-mediated $T_1$ channel} when two devices have the same pointer separation.  {In Fig.~\ref{fig:fig3}, that population-loss channel is evaluated with the full-transverse dressed cavity matrix elements rather than the analytical displaced-RWA Purcell estimate.} It does not predict a single-shot readout fidelity at fixed drive power. A complete readout calculation must solve a driven master equation or quantum-trajectory model that includes physical modulation, finite temperature, measurement-induced transitions, leakage, detector efficiency, resonator nonlinearity, and photon-number constraints.

\bibliographystyle{apsrev4-2}
\bibliography{refs}

\end{document}